\documentclass[showpacs, preprintnumbers, amsmath, amssymb,
  floatfix ,pra, a4paper]{revtex4}
\usepackage[dvips]{graphicx}
\usepackage{bm}
\begin{document} 

\title{Excitations of Bose-Einstein condensates in optical lattices}
\author{K. Braun-Munzinger}
\email{k.braun-munzinger1@physics.ox.ac.uk}
\author{J. A. Dunningham}
\author{K. Burnett}
\affiliation{Clarendon Laboratory, Department of Physics, University of Oxford, Parks Road, Oxford OX1 3PU, United Kingdom.}

\begin{abstract}
In this paper we examine the excitations observable in atoms confined
in an optical lattice around the superfluid-insulator transition. We
use  increases in the number variance of atoms, subsequent to tilting
the lattice as the primary diagnostic of excitations in the lattice.  We show that this
locally determined quantity should be a robust indicator of coherence changes
in the atoms observed in recent experiments. This was found to hold
for commensurate or non-commensurate fillings of the lattice, implying
our results will hold for a wide range of physical cases. Our results
are in good agreement with the quantitative factors of recent
experiments. We do, howevers, find extra features in the excitation
spectra. The variation of the spectra with the duration of the
perturbation also turns out to be an interesting diagnostic of atom dynamics.
\end{abstract}

\maketitle
\section{Introduction}
Interest in the properties of atomic BECs in optical lattices, both theoretical
\cite{sachdev2002,oosten2001,burnett2002} and experimental
\cite{munichtwo,munichone,orzel2001, esslinger03} is growing rapidly at
the moment. This is due
partly to the possibility of studying the physics of a strongly correlated
atomic system for a wide range of the relevant parameters. In particular, the observation of the superfluid (SF) to
Mott insulator (MI) transition has caused a great deal of excitement in future
possibilities for the field. Atomic arrays in optical lattices also promise
a potentially significant route to quantum information processing \cite{jaksch99}. 

The recently introduced experimental technique of adding a linear
potential gradient to the lattice has opened up a new way to study the
phase transition from SF to MI. By carrying out numerical simulations of this
technique we are able to compare the effect of this tilting on states with
various degrees of reduced number fluctuations, i.e. number squeezing. Our focus is on the highly
squeezed regime in the region of the phase transition. 

We will first give a brief overview of the Bose-Hubbard model and the
numerical method we use to simulate the evolution of the trapped condensates in the
presence of the applied field. We then present results of simulations
for a range of initial states, lattice sites and occupation
numbers. In our discussion, we focus on the time dependence of
excitations and the effect of non-commensurate filling. We find that
the variance of the number of atoms at each site is a robust indicator of coherence changes, even
in non-ideal systems (i.e. finite and non-commensurate). Our results
are in excellent qualitative agreement with recent experiments \cite{esslinger03}.
 
\section{Bose-Hubbard model and the energy gap}
The system we simulate in this paper consists of a BEC trapped in a
one-dimensional lattice. We believe, however, that our results have
fairly broad implications for three dimensions as the phenomena we
see are of a generic nature. 
The behaviour of this system is described by the Bose-Hubbard-Model (BHM)
\cite{fisher89,jaksch98}, i.e. 
\begin{equation}
H=\sum_{j}E_{j}\hat{n}_{j}+\frac{1}{2}U\sum_{j}\hat{n}_{j}(\hat{n}%
_{j}-1)-J\sum_{\langle j,k\rangle }\hat{a}_{j}^{\dagger }\hat{a}_{k}.
\end{equation}

Here $\hat{a}_{j},\hat{a}_{j}^{\dagger }$ stand for the bosonic annihilation and
creation operators, $\langle j,k\rangle $ denotes summation over nearest
neighbours, $\hat{n}_{j}=\hat{a}_{j}^{\dagger }\hat{a}_{j}$ is the bosonic number
operator, and $E_{j}$ the energy offset of each site. $J$ is the hopping
matrix element between sites and $U$ the (repulsive) interaction constant
for bosons sharing a site. For numerical simplicity, we will consider
modest-sized lattices with 
between four and eight sites and average occupation numbers of
up to three atoms per site, i.e. $n_{\mathrm{av.occ.}}=1, 2,
3$. Experiments that demonstrate the MI transition involve many more 
lattice sites that this \cite{munichone, munichtwo,
  orzel2001, esslinger03}. However, by considering the evolution of locally determined 
quantities we should be able to gain some insight into the behaviour
of larger lattices. We will see that this is a reasonable assumption
from the fact that the 
results for different numbers of lattice sites show similar
excitation patterns.
We study our system by solving the coupled equations of motion in the number state basis
using an embedded fifth order Runge-Kutta
approximation \cite{numrecipes}. The initial states for the simulations are the
eigenstates of $(1)$ for different values of $U/J$. 
 The squeezed states are then probed by `tilting' the chain of
 sites. In our simulations, this potential gradient is implemented by
adding an onsite energy $E_{j}$ to each site's ground state 
energy. 
The change in the number variance $V$ is then studied after the
potential gradient has been applied for a fixed length of time $\tau
 _{\mathrm{perturb}}$ with instantanuous turn on and turn off. 
We calculate the number variance $V$ as
\begin{equation}
V=\langle (\hat{n}_i)^2\rangle - \langle \hat{n}_i \rangle^2
\end{equation}
where $\hat{n}_i$ is the number operator for site $i$ and
$\langle\rangle$ denotes the ensemble average.
We have chosen to use $V$ as our measure, because it has the great
 advantage of being a locally determined 
quantity, making it less sensitive to the finite size of the
lattices \cite{rey02}. This means that a study of $V$ for
modest lattice sizes should allow us to  
make predictions that are consistent with larger lattices. In the
experiment \cite{munichtwo}, excitations were measured via 
changes in the interference patterns observed in the distribution of
atoms released from the lattice. The excitations, caused by
tilting the lattice, manifest themselves in an increased width of the main interference
peaks \cite{munichtwo} observed when the system is taken back into the
SF phase. In our simulations, the effect of the applied tilt was much clearer in the
 local variable $V$. For further discussion of the use of number
 variance in the theory of BEC and of the relationship between those
 observables see \cite{dunningham98, roth02,rey02, jaksch98}. We should note that
 the number variance may be measured
 experimentally by studying the collapse and revival times of the
 relative phase between sites. The relationship is given by
 \cite{greiner02, wright96}:
\[
\tau_{\mathrm{collapse}}\approx \tau_{\mathrm{revival}}/ \sqrt{V}
\]

One of the main indicators of the MI state is an energy gap. When
present it implies
that atoms can only move between sites if they possess
sufficient energy. When tilting a lattice, one changes the relative potential
energy of the sites and hence the energy available for a hop. Accordingly,
if the energy difference between sites is comparable to the energy gap, boson hopping
should occur. In fact, we expect this to be
 a resonant process when the energy variation
between sites produced by the applied field matches the energy gap. Such
resonant behaviour - which is termed a particle-hole excitation - is
observed both in the experiments \cite{munichone, esslinger03} and the
simulations we present. 
In the SF phase
there is no gap in the excitations and a flow of atoms will occur at all
values of the tilt. 
\begin{figure}[tbp]
\scalebox{0.6}{\includegraphics{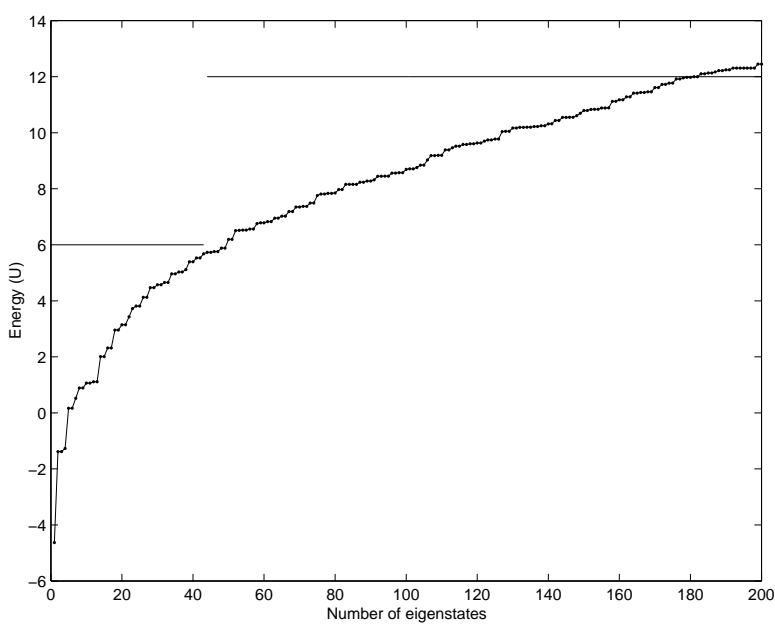}}
\caption{{\protect\footnotesize 
This shows the first 200 eigenstates for U=6 and J = 1 (dots) and U=6 and
J=0.0001 (thin line) 
     }}
\end{figure}
\begin{figure}[tbp]
\scalebox{0.6}{\includegraphics{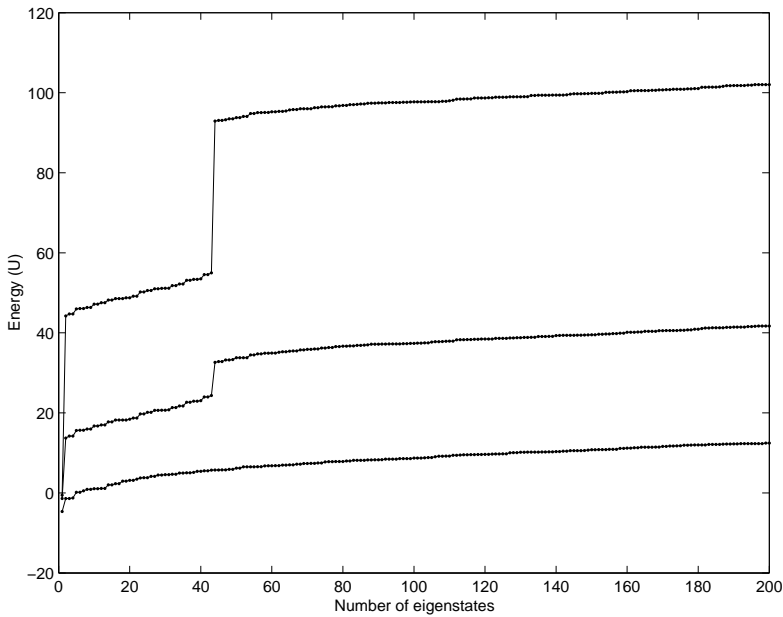}}
\caption{{\protect\footnotesize 
This shows the first 200 eigenstates for U = 6, 20, and 50 and J = 1.  
     }}
\end{figure}
 For an infinite one-dimensional lattice,
mean field and quantum Monte Carlo calculations indicate that the phase transition is
located at $U/J\approx 5.8$ \cite{sheshadri93, freericks95, oosten2001}.
In a finite size lattice, the phase transition is
not sharp and neither is the onset of a gap. In fact, a plot of the eigenstates close to the phase
transition ($U/J=6$) shows a multitude of possible
transitions (see Fig. 1 and 2). Obviously, the simple Mott insulator picture with an
energy gap of $\approx U$ will not explain all excitations possible
for this range of eigenstates. Indeed, we observe a broad smooth
curve superimposed on Mott insulator peaks that slowly vanishes with
increased $U/J$ (see Fig. 6). 
Recent experimental results for lattices with about 100 sites show
very similar features \cite{esslinger03}.  
Fig. 2 shows examples of the eigenstate spectrum for the range of initial states
studied in this paper. Close to the phase transition ($U/J=6$), no gap
is observable. In contrast, states further into the Mott regime
($U/J=20$ and $U/J=50$) show a definite gap, even though the first and
second bands are still broadened.

In the next section, we will present our results before the background
of the theory just discussed. 

\section{Results}

We prepare states with different degrees
of relative number squeezing and then apply a tilt for a perturbation time
$\tau_{\mathrm{perturb}}$.  We then recalculate
$V$ from the resulting wave function. 
This enables us to determine the effect of a fixed tilt on an initial state as a
function of the initial squeezing, the average occupation $n_{\mathrm{av. occ.}}$
and $\tau_{\mathrm{perturb}}$.
In other words, we can study the dependence of excitations on the
perturbation time as well as on the occupation number.
 The results of simulations to that effect can be seen in Fig. 3-7 where 
$V$ is plotted as a function of the applied tilting potential
for various initial states and durations of tilting. 
For a perfectly squeezed state in an infinite lattice, we would expect
resonances at integer multiples and 
integer fractions of the interaction energy $U$ (i.e., at
$U\cdot n_1$ and $U/n_2$, $n_1, n_2$ being
integers, $0<n_1\le$ total number of atoms and  $0<n_2<$ number of
sites).  
It is reasonable to suppose that a resonance at  $U\cdot n_1$ corresponds to $n_1$ particle-hole pairs being
created in adjacent sites while a resonance at
$U/ n_2$ corresponds to the creation of a particle-hole
excitation in two sites $n_2$ sites apart (i.e. in site $i$ and
site $i+n_2$).

Even for a modestly sized lattice (in this case four lattice sites) the
location of peaks was in good agreement with the theoretical
predictions for an infinite lattice. This is clear in Fig. 3,
which shows excitations for filling factors of one, two and three
respectively.

\begin{figure}[tbp]
\scalebox{0.6}{\includegraphics{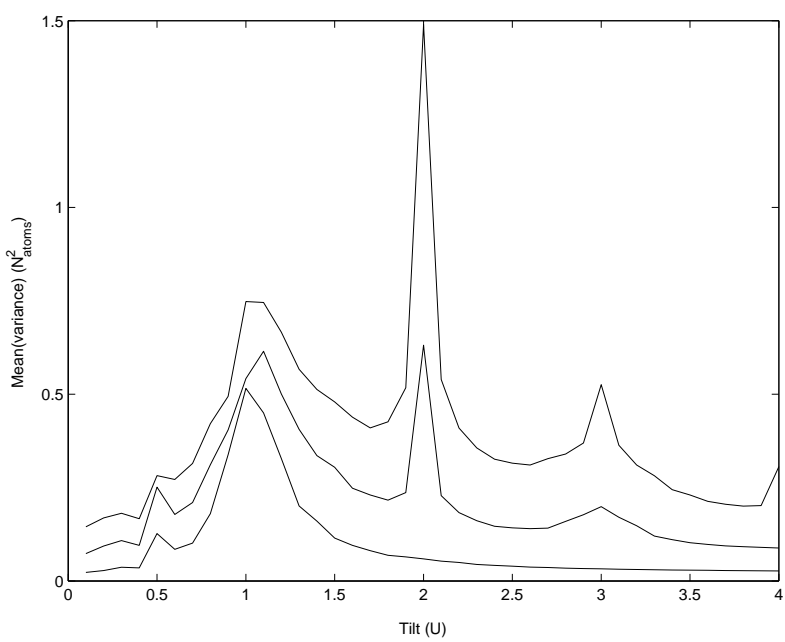}}
\caption{{\protect\footnotesize Excitation pattern for
     $\tau_{\mathrm{perturb}}$ = $2/J$, U=20, J=1, four lattice sites and
     $n_{\mathrm{av. occ.}}=1, 2$ and 3 (bottom to top)
     }}
\end{figure}
 All the plots show one-particle excitations at
$E_{\mathrm{tilt}}\approx 1\cdot U$ and  $E_{\mathrm{tilt}}\approx 0.5\cdot U$. For
$n_{\mathrm{av. occ.}}\le 2$, two, three and even four-particle excitations
appear. Most likely due to the 
superposition of a broad continuous spectrum with discrete peaks,
there is a slight shift ($\approx 0.1-0.2\cdot U$) of the maximum of the resonance at $E_{\mathrm{tilt}}\approx
1\cdot U$ to $E_{\mathrm{tilt}}\approx 1.2\cdot U$ for excitations near
the phase transition, see e.g. Fig. 4. 
\begin{figure}[tbp]
\scalebox{0.6}{\includegraphics{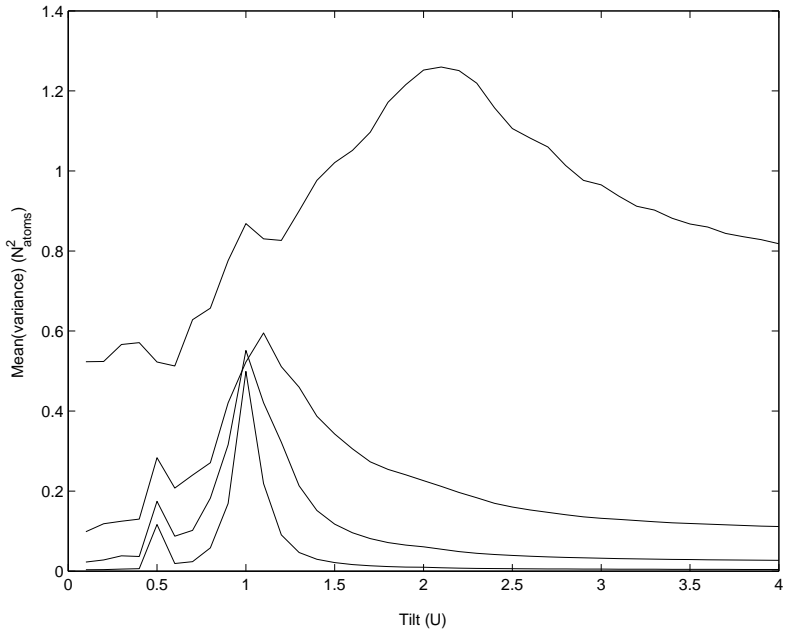}}
\caption{{\protect\footnotesize 
  This plot shows the dependence of excitations on initial
  squeezing. With increasing squeezing (plots are for U = 5, 10, 20
  and 50, J=1 (constant), top to bottom, four lattice sites), resonance effects become noticably narrower and the background decreases.}}
\end{figure}
It is important to
note that a similar effect was
observed in experiments \cite{esslinger03}. 
We found this shift present for all numbers of lattice sites we considered.
The saturation times for lower order processes are of the order of
$1/J$, i.e. the tunneling time (Fig. 5). The first order resonance saturates relatively
quickly, after approximately one tunneling time. The higher order
processes continue to increase in magnitude for a longer period: for
the second order resonance at $2*1/J$, saturation
sets in at about $2*1/J$. Third and fourth order processes take five tunneling times
or longer to saturate.  

\begin{figure}[tbp]
\scalebox{0.7}{\includegraphics{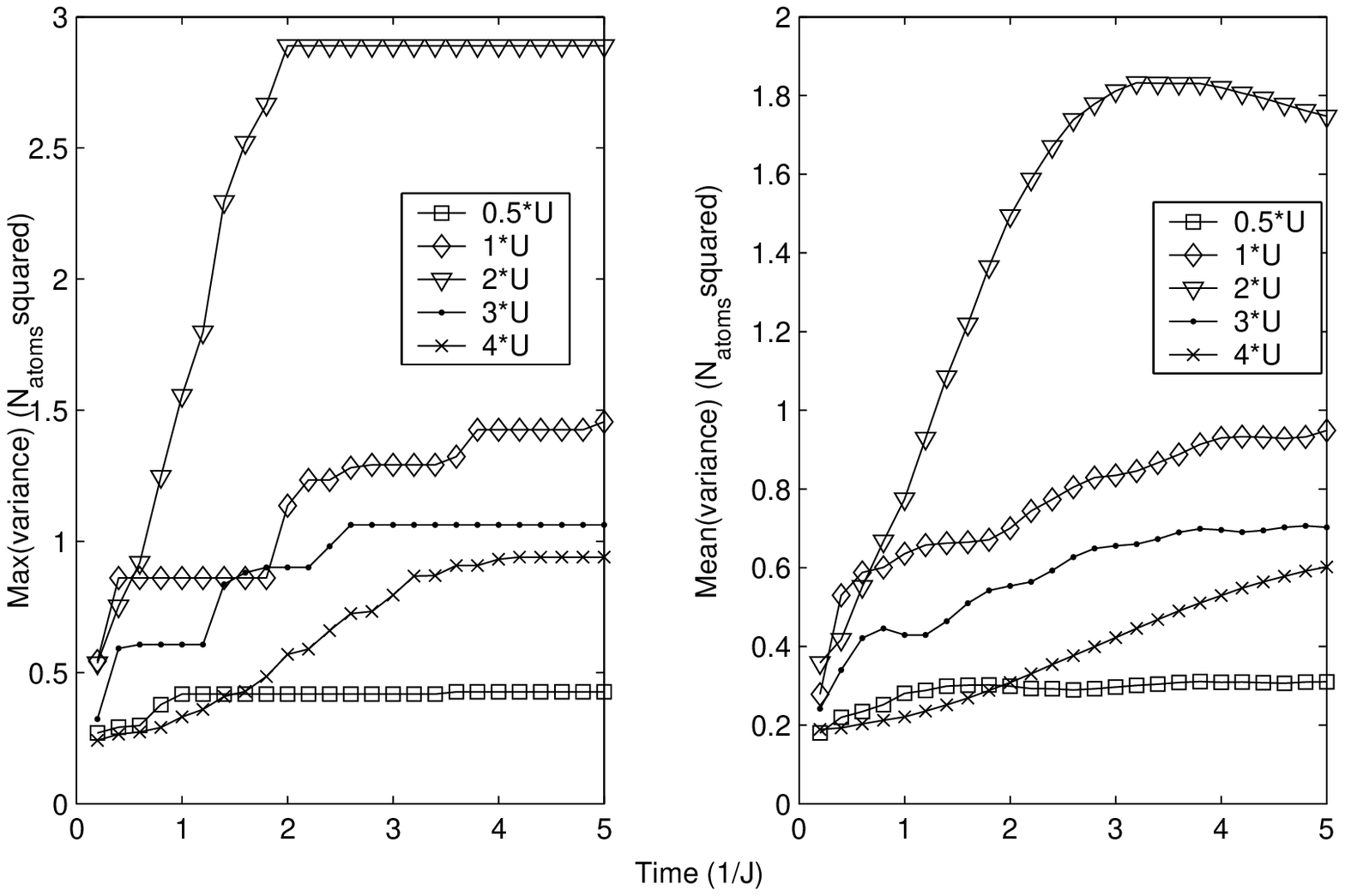}}
\caption{{\protect\footnotesize 
This graph shows the changes in number variance for 0-5 perturbation
times for the five most important peaks ($0.5\cdot U$, $1\cdot U$,
$2\cdot U$, $3\cdot U$ and $4\cdot U$) and $n_{\mathrm{av. occ.}}=3$. The left plot shows the
maximum value of the number variance, the right the average over the
perturbation time.}}
\end{figure}

 With regard to the average number of atoms per well, we
 find that even in a small lattice such as ours, the qualitative
 features are not lost for non-commensurate filling
 factors. As shown in Fig. 6, while non-commensurate filling factors
 result in a more prominent continuous spectrum, they still show discrete
 Mott insulator peaks, albeit with somewhat greater widths . The background level
 is also significantly higher, as to be expected for defects. The
 actual excitations, however, are still clear.
\begin{figure}[tbp]
\scalebox{0.5}{\includegraphics{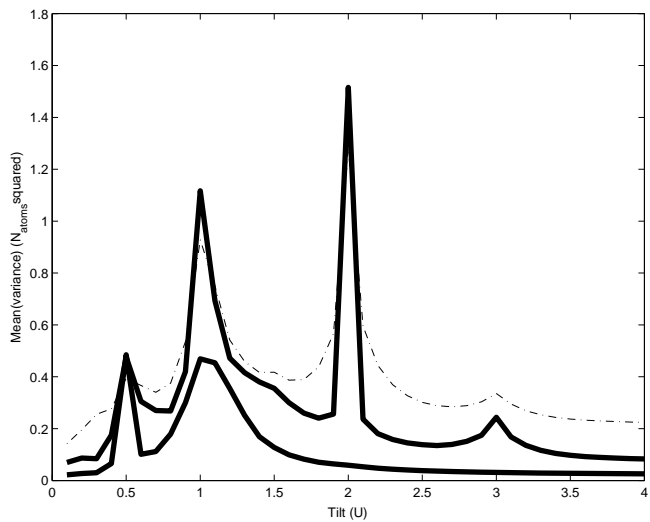}}
\caption{{\protect\footnotesize 
The thick black lines are commensurate (upper $n_{av.occ}=2$, lower
$n_{av. occ.}=1$), while the thin broken line shows results for
$n_{av.occ}=1.8$. There are five lattice sites and $U=20$, $J=1$. }}
\end{figure}

We also compare results for different lattice sizes, ie for four to eight lattice sites.  
For more than five lattice sites and $U/J\ge 20$, the changes in results become very
modest (see Fig. 7). Even for the smaller
configurations, i.e. four and five sites, the important features are still in
very good qualitative agreement. This leads us to be reasonably
confident of the relevance of the principal features of our
calculations for larger realistic systems.

\begin{figure}[tbp]
\scalebox{0.6}{\includegraphics{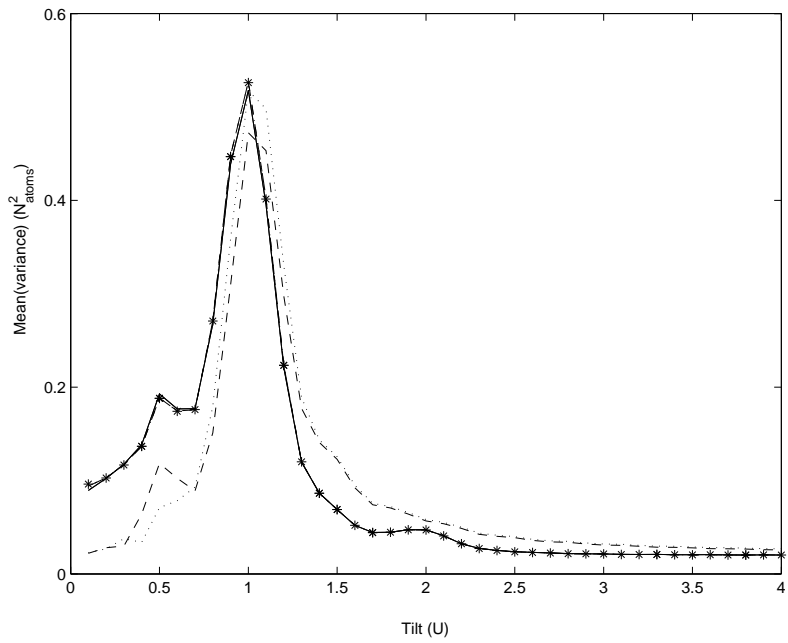}}
\caption{{\protect\footnotesize 
   Results for $U=20$, $J=1$  and $\tau_{\mathrm{perturb}}=1/J$ for four to
   eight lattice sites. Dots: 4 wells, dashes line: 5 wells,
   solid line: 6 wells, dash-dot: 7 wells, stars: 8 wells }}
\end{figure}

\section{Conclusion}
We have shown that the number variance is a indicator for the Mott
insulator gap that is only weakly dependent
on the size of the lattice. Using this diagnostic, we can confirm the
origin of effects
seen in recent experiments \cite{munichtwo, esslinger03}. In addition,
we see higher order effects that fit very well into the
simple picture of excitation of multiples of $U$ in an infinite
lattice. We find that the time dependency of the effects varies with
the complexity of the underlying process: while the simplest
nearest-neighbour hopping from one site to the next saturates in about
a tunneling time, the higher order processes take longer. 
Our results indicate that even non-commensurate filling factors do not
obscure the Mott insulator peaks in the number variance plots. This implies that they could be
useful indicators even in non-ideal systems, such as lattices with defects.

\begin{acknowledgments}

This work was financially supported by the Rhodes Trust, Merton
College, Oxford, the United Kingdom EPSRC, the Royal Society and
Wolfson Foundation and the EU via the network ``Cold Quantum Gases''.

\end{acknowledgments}

\bibliography{basic}

\begin{thebibliography}{18}
\expandafter\ifx\csname natexlab\endcsname\relax\def\natexlab#1{#1}\fi
\expandafter\ifx\csname bibnamefont\endcsname\relax
  \def\bibnamefont#1{#1}\fi
\expandafter\ifx\csname bibfnamefont\endcsname\relax
  \def\bibfnamefont#1{#1}\fi
\expandafter\ifx\csname citenamefont\endcsname\relax
  \def\citenamefont#1{#1}\fi
\expandafter\ifx\csname url\endcsname\relax
  \def\url#1{\texttt{#1}}\fi
\expandafter\ifx\csname urlprefix\endcsname\relax\def\urlprefix{URL }\fi
\providecommand{\bibinfo}[2]{#2}
\providecommand{\eprint}[2][]{\url{#2}}

\bibitem[{\citenamefont{Sachdev et~al.}(2002)\citenamefont{Sachdev, Sengupta,
  and Girvin}}]{sachdev2002}
\bibinfo{author}{\bibfnamefont{S.}~\bibnamefont{Sachdev}},
  \bibinfo{author}{\bibfnamefont{K.}~\bibnamefont{Sengupta}}, \bibnamefont{and}
  \bibinfo{author}{\bibfnamefont{S.}~\bibnamefont{Girvin}},
  \bibinfo{journal}{Phys. Rev. B.} \textbf{\bibinfo{volume}{66}},
  \bibinfo{pages}{075128} (\bibinfo{year}{2002}).

\bibitem[{\citenamefont{van Oosten et~al.}(2001)\citenamefont{van Oosten,
  van~der Straaten, and Stoof}}]{oosten2001}
\bibinfo{author}{\bibfnamefont{D.}~\bibnamefont{van Oosten}},
  \bibinfo{author}{\bibfnamefont{P.}~\bibnamefont{van~der Straaten}},
  \bibnamefont{and} \bibinfo{author}{\bibfnamefont{H.}~\bibnamefont{Stoof}},
  \bibinfo{journal}{Phys. Rev. A} \textbf{\bibinfo{volume}{63}},
  \bibinfo{pages}{053601} (\bibinfo{year}{2001}).

\bibitem[{\citenamefont{Burnett et~al.}(2002)\citenamefont{Burnett, Edwards,
  Clark, and Shotter}}]{burnett2002}
\bibinfo{author}{\bibfnamefont{K.}~\bibnamefont{Burnett}},
  \bibinfo{author}{\bibfnamefont{M.}~\bibnamefont{Edwards}},
  \bibinfo{author}{\bibfnamefont{C.~W.} \bibnamefont{Clark}}, \bibnamefont{and}
  \bibinfo{author}{\bibfnamefont{M.}~\bibnamefont{Shotter}},
  \bibinfo{journal}{J. Phys. B.: At. Mol. Opt. Phys.}
  \textbf{\bibinfo{volume}{35}}, \bibinfo{pages}{1671} (\bibinfo{year}{2002}).

\bibitem[{\citenamefont{Greiner
  et~al.}(2002{\natexlab{a}})\citenamefont{Greiner, Mandel, H\"ansch, and
  Bloch}}]{munichtwo}
\bibinfo{author}{\bibfnamefont{M.}~\bibnamefont{Greiner}},
  \bibinfo{author}{\bibfnamefont{O.}~\bibnamefont{Mandel}},
  \bibinfo{author}{\bibfnamefont{T.~W.} \bibnamefont{H\"ansch}},
  \bibnamefont{and} \bibinfo{author}{\bibfnamefont{I.}~\bibnamefont{Bloch}},
  \bibinfo{journal}{Nature} \textbf{\bibinfo{volume}{419}}, \bibinfo{pages}{51}
  (\bibinfo{year}{2002}{\natexlab{a}}).

\bibitem[{\citenamefont{Greiner
  et~al.}(2002{\natexlab{b}})\citenamefont{Greiner, Mandel, Esslinger,
  H\"ansch, and Bloch}}]{munichone}
\bibinfo{author}{\bibfnamefont{M.}~\bibnamefont{Greiner}},
  \bibinfo{author}{\bibfnamefont{O.}~\bibnamefont{Mandel}},
  \bibinfo{author}{\bibfnamefont{T.}~\bibnamefont{Esslinger}},
  \bibinfo{author}{\bibfnamefont{T.}~\bibnamefont{H\"ansch}}, \bibnamefont{and}
  \bibinfo{author}{\bibfnamefont{I.}~\bibnamefont{Bloch}},
  \bibinfo{journal}{Nature} \textbf{\bibinfo{volume}{415}}, \bibinfo{pages}{39}
  (\bibinfo{year}{2002}{\natexlab{b}}).

\bibitem[{\citenamefont{Orzel et~al.}(2001)\citenamefont{Orzel, Tuchman,
  Fenselau, Yasuda, and Kasevich}}]{orzel2001}
\bibinfo{author}{\bibfnamefont{C.}~\bibnamefont{Orzel}},
  \bibinfo{author}{\bibfnamefont{A.~K.} \bibnamefont{Tuchman}},
  \bibinfo{author}{\bibfnamefont{M.~L.} \bibnamefont{Fenselau}},
  \bibinfo{author}{\bibfnamefont{M.}~\bibnamefont{Yasuda}}, \bibnamefont{and}
  \bibinfo{author}{\bibfnamefont{M.~A.} \bibnamefont{Kasevich}},
  \bibinfo{journal}{Science} \textbf{\bibinfo{volume}{291}},
  \bibinfo{pages}{2386} (\bibinfo{year}{2001}).

\bibitem[{\citenamefont{St\"ofele et~al.}(2003)\citenamefont{St\"ofele, Moritz,
  Schori, K\"ohl, and Esslinger}}]{esslinger03}
\bibinfo{author}{\bibfnamefont{T.}~\bibnamefont{St\"ofele}},
  \bibinfo{author}{\bibfnamefont{H.}~\bibnamefont{Moritz}},
  \bibinfo{author}{\bibfnamefont{C.}~\bibnamefont{Schori}},
  \bibinfo{author}{\bibfnamefont{M.}~\bibnamefont{K\"ohl}}, \bibnamefont{and}
  \bibinfo{author}{\bibfnamefont{T.}~\bibnamefont{Esslinger}}
  (\bibinfo{year}{2003}), \bibinfo{note}{unpublished}.

\bibitem[{\citenamefont{Jaksch et~al.}(1999)\citenamefont{Jaksch, Briegel,
  Cirac, Gardiner, and Zoller}}]{jaksch99}
\bibinfo{author}{\bibfnamefont{D.}~\bibnamefont{Jaksch}},
  \bibinfo{author}{\bibfnamefont{H.-J.} \bibnamefont{Briegel}},
  \bibinfo{author}{\bibfnamefont{J.}~\bibnamefont{Cirac}},
  \bibinfo{author}{\bibfnamefont{C.}~\bibnamefont{Gardiner}}, \bibnamefont{and}
  \bibinfo{author}{\bibfnamefont{P.}~\bibnamefont{Zoller}},
  \bibinfo{journal}{Phys. Rev. Lett.} \textbf{\bibinfo{volume}{82}},
  \bibinfo{pages}{1975} (\bibinfo{year}{1999}).

\bibitem[{\citenamefont{Fisher et~al.}(1989)\citenamefont{Fisher, Weichman,
  G.Grinstein, and Fisher}}]{fisher89}
\bibinfo{author}{\bibfnamefont{M.}~\bibnamefont{Fisher}},
  \bibinfo{author}{\bibfnamefont{B.}~\bibnamefont{Weichman}},
  \bibinfo{author}{\bibnamefont{G.Grinstein}}, \bibnamefont{and}
  \bibinfo{author}{\bibfnamefont{D.}~\bibnamefont{Fisher}},
  \bibinfo{journal}{Phys. Rev. B} \textbf{\bibinfo{volume}{40}},
  \bibinfo{pages}{546} (\bibinfo{year}{1989}).

\bibitem[{\citenamefont{Jaksch et~al.}(1998)\citenamefont{Jaksch, Bruder,
  Cirac, Gardiner, and Zoller}}]{jaksch98}
\bibinfo{author}{\bibfnamefont{D.}~\bibnamefont{Jaksch}},
  \bibinfo{author}{\bibfnamefont{C.}~\bibnamefont{Bruder}},
  \bibinfo{author}{\bibfnamefont{J.}~\bibnamefont{Cirac}},
  \bibinfo{author}{\bibfnamefont{C.}~\bibnamefont{Gardiner}}, \bibnamefont{and}
  \bibinfo{author}{\bibfnamefont{P.}~\bibnamefont{Zoller}},
  \bibinfo{journal}{Phys. Rev. Lett.} \textbf{\bibinfo{volume}{81}},
  \bibinfo{pages}{3108} (\bibinfo{year}{1998}).

\bibitem[{\citenamefont{Press et~al.}(2002)\citenamefont{Press, Teukolsky,
  Vetterling, and Flannery}}]{numrecipes}
\bibinfo{author}{\bibfnamefont{W.}~\bibnamefont{Press}},
  \bibinfo{author}{\bibfnamefont{S.}~\bibnamefont{Teukolsky}},
  \bibinfo{author}{\bibfnamefont{W.}~\bibnamefont{Vetterling}},
  \bibnamefont{and} \bibinfo{author}{\bibfnamefont{B.}~\bibnamefont{Flannery}},
  \emph{\bibinfo{title}{Numerical Recipes in C++, The Art of Scientific
  Computing}} (\bibinfo{publisher}{Cambridge University Press},
  \bibinfo{year}{2002}), chap.~\bibinfo{chapter}{16}.

\bibitem[{\citenamefont{Rey et~al.}(2002)\citenamefont{Rey, Burnett, Roth,
  Edwards, Williams, and Clark}}]{rey02}
\bibinfo{author}{\bibfnamefont{A.}~\bibnamefont{Rey}},
  \bibinfo{author}{\bibfnamefont{K.}~\bibnamefont{Burnett}},
  \bibinfo{author}{\bibfnamefont{R.}~\bibnamefont{Roth}},
  \bibinfo{author}{\bibfnamefont{M.}~\bibnamefont{Edwards}},
  \bibinfo{author}{\bibfnamefont{C.}~\bibnamefont{Williams}}, \bibnamefont{and}
  \bibinfo{author}{\bibfnamefont{C.}~\bibnamefont{Clark}}, \bibinfo{journal}{J.
  Phys. B} \textbf{\bibinfo{volume}{36}}, \bibinfo{pages}{825}
  (\bibinfo{year}{2002}).

\bibitem[{\citenamefont{Dunningham et~al.}(1998)\citenamefont{Dunningham,
  Collett, and Walls}}]{dunningham98}
\bibinfo{author}{\bibfnamefont{J.}~\bibnamefont{Dunningham}},
  \bibinfo{author}{\bibfnamefont{M.}~\bibnamefont{Collett}}, \bibnamefont{and}
  \bibinfo{author}{\bibfnamefont{D.}~\bibnamefont{Walls}},
  \bibinfo{journal}{Phys. Lett. A} \textbf{\bibinfo{volume}{245}},
  \bibinfo{pages}{49} (\bibinfo{year}{1998}).

\bibitem[{\citenamefont{Roth and Burnett}(2002)}]{roth02}
\bibinfo{author}{\bibfnamefont{R.}~\bibnamefont{Roth}} \bibnamefont{and}
  \bibinfo{author}{\bibfnamefont{K.}~\bibnamefont{Burnett}}
  (\bibinfo{year}{2002}), \eprint{cond-mat/0209066}.

\bibitem[{\citenamefont{Greiner
  et~al.}(2002{\natexlab{c}})\citenamefont{Greiner, Mandel, and
  H\"ansch}}]{greiner02}
\bibinfo{author}{\bibfnamefont{M.}~\bibnamefont{Greiner}},
  \bibinfo{author}{\bibfnamefont{O.}~\bibnamefont{Mandel}}, \bibnamefont{and}
  \bibinfo{author}{\bibfnamefont{T.~W.} \bibnamefont{H\"ansch}},
  \bibinfo{journal}{Nature} \textbf{\bibinfo{volume}{419}}, \bibinfo{pages}{51}
  (\bibinfo{year}{2002}{\natexlab{c}}).

\bibitem[{\citenamefont{Wright et~al.}(1996)\citenamefont{Wright, Walls, and
  H\"ansch}}]{wright96}
\bibinfo{author}{\bibfnamefont{E.~M.} \bibnamefont{Wright}},
  \bibinfo{author}{\bibfnamefont{D.}~\bibnamefont{Walls}}, \bibnamefont{and}
  \bibinfo{author}{\bibfnamefont{T.~W.} \bibnamefont{H\"ansch}},
  \bibinfo{journal}{Phys. Rev. Lett.} \textbf{\bibinfo{volume}{77}},
  \bibinfo{pages}{2158} (\bibinfo{year}{1996}).

\bibitem[{\citenamefont{Sheshadri et~al.}(1993)\citenamefont{Sheshadri,
  Krishnamurthy, Pandit, and Ramakrishnan}}]{sheshadri93}
\bibinfo{author}{\bibfnamefont{K.}~\bibnamefont{Sheshadri}},
  \bibinfo{author}{\bibfnamefont{H.~R.} \bibnamefont{Krishnamurthy}},
  \bibinfo{author}{\bibfnamefont{R.}~\bibnamefont{Pandit}}, \bibnamefont{and}
  \bibinfo{author}{\bibfnamefont{R.}~\bibnamefont{Ramakrishnan}},
  \bibinfo{journal}{Europhys. Lett.} \textbf{\bibinfo{volume}{22}},
  \bibinfo{pages}{257} (\bibinfo{year}{1993}).

\bibitem[{\citenamefont{Freericks and Monien}(1995)}]{freericks95}
\bibinfo{author}{\bibfnamefont{J.}~\bibnamefont{Freericks}} \bibnamefont{and}
  \bibinfo{author}{\bibfnamefont{H.}~\bibnamefont{Monien}},
  \bibinfo{journal}{Europhy. Lett.} \textbf{\bibinfo{volume}{26}},
  \bibinfo{pages}{545} (\bibinfo{year}{1995}).

\end{thebibliography}

\end{document}